\documentclass[runningheads]{llncs}
\usepackage[T1]{fontenc}
\usepackage{graphicx}
\usepackage{caption}
\usepackage{subcaption}
\usepackage{graphicx}
\usepackage{booktabs}       
\usepackage{amsfonts}       
\usepackage{nicefrac}       
\usepackage{microtype}      
\usepackage{xcolor}         
\usepackage{amsmath}
\usepackage[misc]{ifsym}
\usepackage{algorithm}
\usepackage{algcompatible}
\usepackage{algpseudocode}
\begin{document}
%
\title{Uncovering Privacy Vulnerabilities through Analytical Gradient Inversion Attacks}
%
%
\author{Tamer Ahmed Eltaras\inst{1}\textsuperscript{\Letter}\orcidID{0000-0002-8664-9091} \and
Qutaibah Malluhi\inst{2,3}\orcidID{0000-0003-2849-0569} \and
Alessandro Savino\inst{3}\orcidID{0000-0003-0529-7950}\and
Stefano Di Carlo\inst{3}\orcidID{0000-0002-7512-5356}\and
Adnan Qayyum\inst{3}\orcidID{2222--3333-4444-5555}}
%
%
\institute{Politecnico di Torino, Control and Computer Engineering Department, Turin 10129, Italy \and
Qatar University, Doha, Qatar \and
Information Technology University of the Punjab, Lahore, Pakistan
}
\maketitle              
\begin{abstract}

Federated learning has emerged as a prominent privacy-preserving technique for leveraging large-scale distributed datasets by sharing gradients instead of raw data. However, recent studies indicate that private training data can still be exposed through gradient inversion attacks. While earlier analytical methods have demonstrated success in reconstructing input data from fully connected layers, their effectiveness significantly diminishes when applied to convolutional layers, high-dimensional inputs, and scenarios involving multiple training examples. This paper extends our previous work \cite{eltaras2024r} and proposes three advanced algorithms to broaden the applicability of gradient inversion attacks. The first algorithm presents a novel data leakage method that efficiently exploits convolutional layer gradients, demonstrating that even with non-fully invertible activation functions, such as ReLU, training samples can be analytically reconstructed directly from gradients without the need to reconstruct intermediate layer outputs. Building on this foundation, the second algorithm extends this analytical approach to support high-dimensional input data, substantially enhancing its utility across complex real-world datasets. The third algorithm introduces an innovative analytical method for reconstructing mini-batches, addressing a critical gap in current research that predominantly focuses on reconstructing only a single training example. Unlike previous studies that focused mainly on the weight constraints of convolutional layers, our approach emphasizes the pivotal role of gradient constraints, revealing that successful attacks can be executed with fewer than 5\% of the constraints previously deemed necessary in certain layers. 
\keywords{Gradient Inversion Attacks  \and Data Leakage \and Federated Learning.}
\end{abstract}

\section{Introduction}
\label{introduction}
In recent years, the widespread adoption of machine learning (ML) and deep learning (DL) models across various domains has necessitated the utilization of high-quality, large-scale datasets. However, this demand has introduced critical privacy concerns, particularly when handling sensitive data such as medical records, financial transactions, and personal communications. Conventional centralized learning approaches that require aggregating data on the central server for model training pose significant privacy and data breach risks. To address these challenges, Federated Learning (FL) \cite{mcmahan2017communication} has emerged as a promising paradigm that reconciles the need for collaborative model training with stringent privacy requirements. FL enables multiple clients to collaboratively train a global model without exposing their raw data, thereby enhancing data privacy. In this framework, clients iteratively train the model locally using their private data and transmit only the computed gradients to a central server for aggregation and global model updates. However, recent studies have revealed vulnerabilities in the FL's gradient-sharing mechanism. Specifically, it has been demonstrated that shared gradients can be exploited to reconstruct sensitive training data, thereby challenging the presumed privacy of FL ~\cite{zhao2020idlg,zhu2019deep,wang2019beyond,wei2020framework,geiping2020inverting,Sabry2023WearableDG,eltaras2019partial}. These gradient inversion attacks represent a significant threat, especially in privacy-sensitive applications such as healthcare and finance, where data confidentiality is critical.



The threat posed by gradient inversion attacks has become a critical area of research in FL security. For instance, early works such as Deep Leakage from Gradients (DLG) \cite{zhu2019deep} and its enhanced version iDLG \cite{zhao2020idlg} had demonstrated the feasibility of reconstructing training data by optimizing random inputs until their gradient signatures matches the observed gradients. While these methods established the practicality of gradient inversion attacks, they often suffer from high computational overhead and limited scalability. To address these challenges, Zhu et al. \cite{zhu2020r} proposed an analytical method known as R-GAP, which formulates closed-form solutions for data reconstruction. Although R-GAP represents a significant advancement by reducing computational costs, it primarily focuses on fully connected layers and assumes the use of fully invertible activation functions. However, convolutional layers, which are integral to modern neural networks, introduce unique structural complexities such as local receptive fields and parameter-sharing mechanisms—factors that R-GAP fails to adequately address. Furthermore, the widespread adoption of non-fully invertible activation functions like ReLU in Convolutional Neural Networks (CNNs) further complicates the gradient inversion process. This gap in existing literature highlights the need for more robust analytical methods capable of effectively exploiting convolutional gradients for data reconstruction, even in the presence of non-invertible activation functions.


In this paper, we build upon our previous work \cite{eltaras2024r}, where we introduced R-CONV, a novel analytical method designed to reconstruct training data from the gradients of convolutional layers.
 Our contributions in this work are fourfold:
\begin{enumerate}

    \item Extension of R-CONV to High-Dimensional Data Reconstruction: We extend the original R-CONV method to support the reconstruction of high-dimensional input data from convolutional layer gradients, significantly broadening its applicability.. 

    \item Analytical Reconstruction in Mini-Batch Settings: We introduce a novel analytical algorithm capable of reconstructing data from mini-batches, addressing a critical gap in existing research, which has focused primarily on single-example reconstruction. To the best of our knowledge, this is the first work to analytically demonstrate the feasibility of gradient inversion attacks in mini-batch settings.

    \item Gradient Constraints vs. Weight Constraints: We highlight the critical role of gradient constraints over traditional weight constraints in convolutional layers. Our findings reveal that in some network layers, gradient constraints can expose significantly more information about the input data than weight constraints.

    \item Influence of Convolutional Parameters on Attack Effectiveness: We conduct a systematic in-depth analysis of how variations in convolutional layer parameters—such as kernel size, stride, and padding—affect the effectiveness of gradient-based attack techniques. 

\end{enumerate}

\section{Related Work}
\label{sec:related}
In the literature, it has been demonstrated that shared model updates in FL can inadvertently leak private information. For instance, early works in this line of research focused on ``shallow'' gradient leakage attacks such as membership inference attacks, where attackers determine if a particular data sample was used during training \cite{melis2019exploiting},  and on using generative models to produce samples similar to the training data \cite{hitaj2017deep}. These initial investigations not only highlighted the privacy risks in FL but also provided the foundation for more direct approaches for reconstructing training data from shared gradients, such as Deep Gradient Leakage (DLG) \cite{zhu2019deep}. Below, we provide an overview of similar studies, categorizing them into two categories: optimization-based and analytical approaches.


\subsection{Optimization-Based Approaches} 

DLG is the first method for directly recovering training data from shared gradients, as proposed by Zhu et al. \cite{zhu2019deep}. DLG is an optimization-based algorithm that reconstructs training data by leveraging ``dummy'' gradients and their corresponding class labels. Instead of updating model weights, this approach minimizes the discrepancy between these dummy gradients—computed from random inputs—and the actual gradients obtained during training. This process enables the recovery of the original inputs, with experiments showing accurate pixel-level reconstruction for batch sizes of up to eight samples. However, this technique is largely limited to simple neural network architectures that use sigmoid activation functions and is primarily applicable to low-dimensional datasets, such as the CIFAR images \cite{krizhevsky2009learning}. As a result, it does not extend well to high-dimensional images or models with more complex architectures that employ non-fully invertible activation functions like ReLU. Building on this formulation, Geiping et al. \cite{geiping2020inverting} refined the approach by incorporating an additional regularization term into the optimization process. They replaced the Euclidean distance with cosine similarity for measuring reconstruction loss and introduced total variation regularization to enhance the stability and performance of the reconstruction. These modifications allowed their method to effectively recover low-dimensional images for batch sizes of up to one hundred samples. Subsequent studies have attempted to improve the performance of gradient inversion attacks in more realistic settings. For example, Zhao et al. \cite{zhao2020idlg} improved on DLG by consistently inferring reference labels directly from the gradients, thus enhancing the precision of the reconstructed data.

\subsection{Analytical Approaches} 
In addition to optimization-based methods, several analytical techniques have been introduced in the literature that provide a theoretical understanding of gradient inversion attacks. For instance, Zhu et al. \cite{zhu2020r} proposed a closed-form iterative method, named R-GAP, which not only reconstructs training data from gradients but also employs a rank analysis to assess the susceptibility of different network architectures to these attacks. Building on this work, Chen et al. \cite{chen2021understanding} introduced the Combined Optimization Attack (COPA), reformulating the reconstruction objective in order to explicitly capture the constraints that gradient information imposes on the inversion process. In a recent study, Zhang et al. \cite{zhang2024understanding} developed an analytical tool based on the Inversion Influence Function (I2F), which leverages gradients and Jacobian-vector products to examine how perturbations affect the reconstruction process. Their approach emphasizes the influence of network initialization and the structure of the Jacobian on privacy leakage. Similarly, Wang et al. \cite{wang2023reconstructing} demonstrated that it is feasible to fully reconstruct training data from a single gradient query under certain conditions. A notable limitation of the aforementioned analytical methods is their reliance on the assumption of fully invertible activation functions. Our work addresses this gap by extending these analytical insights to convolutional layers in CNNs. 
Unlike previous methods that primarily rely on weight constraints and assume invertible activations, our approach focuses on exploiting gradient constraints. This focus enables effective data reconstruction even when non-invertible activation functions are employed, thereby overcoming a key limitation in existing analytical approaches.

\section{Threat Model}
This work explores gradient inversion attacks within the framework of horizontal federated learning (FL), where participating clients collaboratively train a shared model using their local datasets while keeping raw data private. The target model is a neural network architecture comprising convolutional layers followed by fully connected layers, a common configuration in domains such as image processing, natural language processing, and other data-driven applications. Our threat model assumes an honest-but-curious passive adversary who adheres to the FL protocol and does not alter model parameters or deviate from established rules. Unlike an active adversary that might inject malicious updates or manipulate model weights, this passive attacker remains constrained by the protocol and operates only within the system’s legitimate operations. The adversary is assumed to know the model architecture and initial parameters and to have access to the gradient updates exchanged during the FL process since such information is typically shared with the server. However, the adversary lacks any direct access to the clients’ local data and has no prior knowledge of the underlying data distribution. The adversary’s primary objective is to exploit the received gradient updates, in conjunction with their knowledge of the model parameters, to reconstruct the clients’ original input data, thereby exposing potential privacy vulnerabilities in federated learning systems.
\section{Methodology}
\label{sec:Methodlogy}
In this section, we outline our methodology for reconstructing input examples by leveraging both gradient updates and initial model parameters. We introduce three algorithms that address different aspects of gradient inversion attacks:

\begin{enumerate}

    \item \textbf{Single training example reconstruction} by utilizing gradient updates even when non-linear activation functions are employed.
    \item \textbf{High-resolution image reconstruction} extending the previous approach.
    \item \textbf{Reconstruction of images from mini-batches}, capturing multiple training samples simultaneously.

\end{enumerate}

The target model is a neural network architecture comprising convolutional layers followed by fully connected layers. The following notation is used throughout the paper:

\begin{itemize}
    \item\textbf{$W^{(i)}$, $b^{(i)}$:} The initial weight and bias parameters associated with layer $i$, where $i$ is an index from 1 to $d$, and $d$ is the total count of layers.
    \item\textbf{$\nabla W^{(i)}\ell$, $\nabla b^{(i)}\ell$:} The gradients of the loss function \textbf{$\ell$} with respect to the weights and biases at layer $i$.
    \item\textbf{$ X^{(i)}$:} Input received by layer $i$.
    \item\textbf{$ A^{(i)}$:} The activation function of layer $i$.
    \item\textbf{$ Z^{(i)}$:} The output of layer $i$ before applying the activation function.
    \item\textbf{$ O^{(i)}$:} The final output of layer $i$ after the activation function is applied, which is equivalent to $ X^{(i+1)}$. 
    \item\textbf{$ N^{(i)}$:} Nodes in fully connected layer $i$ that serve as upstream connections.
    \item\textbf{$ C^{(i)}$:} Nodes in fully connected layer $i$ that serve as downstream connections; in the final fully connected layer $d$, this represents the target classes.

\end{itemize}

\subsection{Single Training Example Reconstruction}
We propose a method for reconstructing a single training example that overcomes the limitations of existing methods \cite{zhu2020r,chen2021understanding}, which are often restricted to networks employing fully-invertible activation functions (e.g., Leaky ReLU).
By focusing on reconstructing the gradient with respect to the output, rather than the output itself, our approach leverages a key property found in most common activation functions: the derivative of an activation function’s output can be expressed in terms of its output. This insight allows our method to bypass the invertibility constraint.
\newline
The reconstruction process follows three primary steps:
\begin{itemize}
    \item \textbf{Constructing the gradient of the input for the first fully connected layer:} The process begins by establishing the gradient information related to the input at this layer.
    \item \textbf{Backpropagating the gradient through the activation function to the previous convolutional layer:} This step allows the extraction of gradient information relevant to the outputs of convolutional layers.
    \item \textbf{Reconstructing input and activation gradients for convolutional layers:} This enables further propagation of gradients to reconstruct the input at each preceding convolutional layer iteratively. The process continues until reaching the first layer, where the reconstructed input corresponds to the original training example.
\end{itemize}
The following sections provide a step-by-step explanation of this approach, followed by a formalized algorithm in Algorithm \ref{alg:single_example_reconstruction}.

\subsubsection{Constructing the gradient of the input for the first fully connected layer:}
Let \( f \) denote the index of the first fully connected layer in the network architecture, with the input vector represented as
\(
X^{(f)} = (x_1^{(f)}, x_2^{(f)}, \dots, x_{N^{(f)}}^{(f)}),
\) where \( N^{(f)} \) is the number of upstream (input) nodes in this layer. The number of downstream (output) nodes in this layer is denoted by \( C^{(f)} \). The network’s loss function is expressed as \(
\ell = \ell(f(X^d), y),
\) where \( d \) represents the index of the final layer, and \( y \) is the true output label. In this fully connected layer, the input vector \( X^{(f)} \) is processed by a weight matrix \( W^{(f)} \in \mathbb{R}^{C^{(f)} \times N^{(f)}} \) and a bias vector \(
b^{(f)} = (b_1^{(f)}, b_2^{(f)}, \dots, b_{C^{(f)}}^{(f)}),
\) to generate an intermediate output \( Z^{(f)} \). The weight matrix connects upstream nodes to downstream nodes, with each element \( W_{ij}^{(f)} \) representing the weight between the \( j \)-th input node and the \( i \)-th output node. The activation function \( A^{(f)} \) is applied to the intermediate output \( Z^{(f)} \) to produce the final output of the layer, \( O^{(f)} \), which also serves as the input to the subsequent layer \( (f+1) \).
\newline
The objective is to compute the gradient of the loss function \( \ell \) with respect to the input vector \( X^{(f)} \) of the first fully connected layer. This process involves systematically applying the chain rule to effectively propagate gradients from the loss function back to the input through the intermediate outputs and weights.
\newline
The intermediate output \( Z^{(f)} \) of the fully connected layer is computed using the linear transformation:

\begin{equation}
Z^{(f)} = W^{(f)} X^{(f)} + b^{(f)} \tag{1}
\end{equation}

To compute the gradient of the loss function \( \ell \) with respect to the input vector \( X^{(f)} \), the chain rule is applied as follows:

\begin{equation}
\nabla_{X^{(f)}} \ell = \frac{\partial \ell}{\partial X^{(f)}} = \frac{\partial \ell}{\partial Z^{(f)}} \cdot \frac{\partial Z^{(f)}}{\partial X^{(f)}} \tag{2}
\end{equation}

The gradient of the intermediate output \( Z^{(f)} \) with respect to the input \( X^{(f)} \) can be directly derived from Equation (1). Given that \( Z^{(f)} \) is a linear transformation of \( X^{(f)} \):

\begin{equation}
\frac{\partial Z^{(f)}}{\partial X^{(f)}} = W^{(f)} \tag{3}
\end{equation}

For the bias term \( b^{(f)} \), the gradient of the loss function is computed using the chain rule:

\begin{equation}
\nabla_{b^{(f)}} \ell = \frac{\partial \ell}{\partial b^{(f)}} = \frac{\partial \ell}{\partial Z^{(f)}} \cdot \frac{\partial Z^{(f)}}{\partial b^{(f)}} \tag{4}
\end{equation}

From Equation (1), the partial derivative of the intermediate output with respect to the bias is:

\[
\frac{\partial Z^{(f)}}{\partial b^{(f)}} = 1
\]

Thus, the gradient of the loss function with respect to the bias simplifies to:

\begin{equation}
\nabla_{b^{(f)}} \ell = \frac{\partial \ell}{\partial b^{(f)}} = \frac{\partial \ell}{\partial Z^{(f)}} \tag{5}
\end{equation}

By combining Equations (2), (3), and (5), the gradient of the loss function with respect to the input vector \( X^{(f)} \) is obtained as:

\begin{equation}
\nabla_{X^{(f)}} \ell = W^{(f)T} \cdot \nabla_{b^{(f)}} \ell \tag{6}
\end{equation}

where \( W^{(f)T} \) denotes the transpose of the weight matrix \( W^{(f)} \). Since both \( W^{(f)T} \) and \( \nabla_{b^{(f)}} \ell \) are known, the gradient \( \nabla_{X^{(f)}} \ell\) can be computed directly.
\subsubsection{Backpropagating the gradient through the activation function to the previous convolutional layer:}

After computing the input gradient \( \nabla_{X^{(f)}} \ell \) of the fully connected layer \( f \), the next step involves backpropagating this gradient through the activation function of the preceding convolutional layer \( (f-1) \). For the convolutional layer \( (f-1) \), let \( Z^{(f-1)} \) denote the intermediate output of the layer before applying the activation function, and let \( O^{(f-1)} \) represent the output of the layer after applying the activation function. Notably, \( O^{(f-1)} \) also serves as the input to the subsequent layer $X^{(f)}$.
So, the relationship between these variables through the activation function \( A^{(f-1)} \) is:

\[
O^{(f-1)} = A^{(f-1)}(Z^{(f-1)}) \quad \text{and} \quad O^{(f-1)} = X^{(f)} \quad \text{with} \quad \nabla_{O^{(f-1)}} \ell = \nabla_{X^{(f)}} \ell
\]

To compute the gradient of the loss function \( \ell \) with respect to the intermediate output \( Z^{(f-1)} \), we apply the chain rule:

\[
\nabla_{Z^{(f-1)}} \ell = \frac{\partial \ell}{\partial Z^{(f-1)}} = \frac{\partial \ell}{\partial O^{(f-1)}} \cdot \frac{\partial O^{(f-1)}}{\partial Z^{(f-1)}}
\]

Thus, the gradient can be reformulated as:

\[
\nabla_{Z^{(f-1)}} \ell = \nabla_{X^{(f)}} \ell \cdot A'^{(f-1)}(Z^{(f-1)}) \tag{8}
\]

where \( A'^{(f-1)} \) is the derivative of the activation function at layer \( (f-1) \) with respect to its input \( Z^{(f-1)} \).
\newline
The derivatives of most activation functions can be expressed solely in terms of their outputs. Table~\ref{tab:activation_derivatives} presents the derivatives of commonly used activation functions, reformulated in terms of the input vector \( X^{(f)} \).

\begin{table}[h]
\centering
\caption{Derivative of Activation Functions}
\label{tab:activation_derivatives}
\setlength{\tabcolsep}{8.5pt} 
\renewcommand{\arraystretch}{2.2} 
\begin{tabular}{|l|l|l|}
\hline
\textbf{Name} & \textbf{Equation} & \textbf{Derivative Expressed in \( X^{(f)} \)} \\ \hline
Sigmoid & 
\( A^{(f-1)}(Z^{(f-1)}) = \frac{1}{1+e^{-Z^{(f-1)}}} \) & 
\( A'^{(f-1)}(Z^{(f-1)}) = X^{(f)}(1-X^{(f)}) \) \\ \hline

Tanh & 
\( A^{(f-1)}(Z^{(f-1)}) = \tanh(Z^{(f-1)}) \) & 
\( A'^{(f-1)}(Z^{(f-1)}) = 1 - (X^{(f)})^2 \) \\ \hline

ArcTan & 
\( A^{(f-1)}(Z^{(f-1)}) = \tan^{-1}(Z^{(f-1)}) \) & 
\( A'^{(f-1)}(Z^{(f-1)}) = \frac{1}{1 + \tan(X^{(f)})^2} \) \\ \hline

SoftPlus & 
\( A^{(f-1)}(Z^{(f-1)}) = \log_e(1+e^{Z^{(f-1)}}) \) & 
\( A'^{(f-1)}(Z^{(f-1)}) = \frac{1}{1+e^{-Z^{(f-1)}}} \) \\ \hline

ReLU & 
\( A^{(f-1)}(Z^{(f-1)}) = \begin{cases} 
0 & Z^{(f-1)} \leq 0 \\ 
Z^{(f-1)} & Z^{(f-1)} > 0 
\end{cases} \) & 
\( A'^{(f-1)}(Z^{(f-1)}) = \begin{cases} 
0 & X^{(f)} \leq 0 \\ 
1 & X^{(f)} > 0 
\end{cases} \) \\ \hline

Leaky ReLU & 
\( A^{(f-1)}(Z^{(f-1)}) = \begin{cases} 
\alpha Z^{(f-1)} & Z^{(f-1)} < 0 \\ 
Z^{(f-1)} & Z^{(f-1)} \geq 0 
\end{cases} \) & 
\( A'^{(f-1)}(Z^{(f-1)}) = \begin{cases} 
\alpha & X^{(f)} < 0 \\ 
1 & X^{(f)} \geq 0 
\end{cases} \) \\ \hline
\end{tabular}
\end{table}

As demonstrated in \cite{aono2017privacy}, the input vector \( X^{(f)} \) of the fully connected layer \( f \) can be computed using the gradients of the weights and biases of the layer. The input vector \( X^{(f)} \) is derived as:

\[
X^{(f)T} = \frac{\partial \ell}{\partial W^{(f)}_j} \bigg/ \frac{\partial \ell}{\partial b^{(f)}_j} \tag{9}
\]

where \( j \) denotes the \( j \)-th row of the weight matrix \( W^{(f)} \). This approach enables the reconstruction of the input vector using only the available gradient information.
\newline

By combining the computed input vector \( X^{(f)} \) with the known activation function derivatives and substituting it into Equation (8), the output gradient of the preceding convolutional layer before applying the activation function can be reconstructed accurately. 


\subsubsection{Reconstructing input and activation gradients for convolutional layers::}

We now use the computed gradient \( \nabla Z^{(f-1)}\ell \) to achieve two main objectives:
\begin{itemize}

    \item Compute the gradient with respect to the input of the layer \( (f-1) \)
    \item Reconstruct the input of the layer \( (f-1) \)
\end{itemize}

The gradient with respect to the input can be computed using:

\begin{equation}
\frac{\partial L}{\partial X^{(f-1)}} = W^{(f-1)}_{\text{flipped}} * \frac{\partial L}{\partial Z^{(f-1)}} \tag{10}
\end{equation}

where:
\begin{itemize}
    \item \( W^{(f-1)}_{\text{flipped}} \) is the convolutional kernel \( W^{(f-1)} \) flipped both horizontally and vertically.
    \item The flipping operation ensures correct alignment of the gradients during the convolution.
    \item \( * \) denotes the convolution operation.
\end{itemize}

To reconstruct the input \( X^{(f-1)} \) of a convolutional layer, we utilize the relationship between the convolutional weights, the input, and the output gradients. The convolutional layer's output before the activation function is:

\begin{equation}
Z^{(f-1)} = W^{(f-1)} * X^{(f-1)} + b^{(f-1)} \tag{11}
\end{equation}

The gradient of the loss function \( \ell \) with respect to the convolutional weights \( W^{(f-1)} \) is computed using:

\begin{equation}
\frac{\partial \ell}{\partial W^{(f-1)}} = \frac{\partial \ell}{\partial Z^{(f-1)}} * X^{(f-1)} \tag{12}
\end{equation}

This formula shows that the gradient with respect to the weights is the convolution of the input \( X^{(f-1)} \) with the gradient of the output \( \frac{\partial \ell}{\partial Z^{(f-1)}} \). The convolution operation involves sliding the output gradient over the input and accumulating the results, effectively capturing the sensitivity of each weight in the filter to changes in the input.

To reconstruct the input, this relationship can be expressed as a linear system of equations. For each weight \( W_{i,j}^{(f-1)} \), the gradient is:

\begin{equation}
\frac{\partial \ell}{\partial W_{i,j}^{(f-1)}} = \sum_m \frac{\partial \ell}{\partial Z_m^{(f-1)}} \cdot X^{(f-1)}[t[m]] \tag{13}
\end{equation}

where:
\begin{itemize}
    \item \( \frac{\partial \ell}{\partial W_{i,j}^{(f-1)}} \) is the gradient with respect to the convolutional weight.
    \item \( \frac{\partial \ell}{\partial Z_m^{(f-1)}} \) represents the gradient with respect to the output element before activation.
    \item \( X^{(f-1)}[t[m]] \) are the input elements contributing to the output through the convolution operation.
    \item \( t[m] \) is the set of indices indicating the receptive field of the convolution at position \( m \).
\end{itemize}

The resulting linear system can be represented in matrix form as:

\begin{equation}
\mathbf{\nabla W}^{(f-1)} = \mathbf{\nabla Z}^{(f-1)} * \mathbf{X}^{(f-1)} \tag{14}
\end{equation}

where:
\begin{itemize}
    \item \( \mathbf{\nabla W}^{(f-1)} = \frac{\partial \ell}{\partial W^{(f-1)}} \) is the matrix of gradients with respect to the convolutional weights in layer \( (f-1) \).
    \item \( \mathbf{\nabla Z}^{(f-1)} = \nabla_{Z^{(f-1)}} \ell \) is the matrix of output gradients before the activation function at layer \( (f-1) \).
    \item \( \mathbf{X}^{(f-1)} = X^{(f-1)} \) is the input matrix to be reconstructed for the layer \( (f-1) \).
\end{itemize}

The input \( X^{(f-1)} \) can be reconstructed by solving the linear system:

\begin{equation}
\mathbf{X}^{(f-1)} = \mathbf{\nabla Z}^{(f-1)^{-1}} * \mathbf{\nabla W}^{(f-1)} \tag{15}
\end{equation}

where \( \mathbf{\nabla Z}^{(f-1)^{-1}} \) is the inverse of the gradient matrix \( \mathbf{\nabla Z}^{(f-1)} \). In the following section, we will analyze when this reconstruction is possible, depending on the number of convolutional filters. The feasibility of solving this linear system relies on whether the number of available convolutional filters provides a sufficient number of equations to determine all unknowns in the input matrix.

If the current convolutional layer is not the first layer in the network, we can repeat this process. By propagating the gradients to the preceding layers, we iteratively compute both the input gradients and the input feature maps of each earlier layer. This iterative approach continues until reaching the first layer, where the reconstructed input corresponds to the original input data required by the network.
\begin{algorithm}[t]
\caption{Single Training Example Reconstruction}
\label{alg:single_example_reconstruction}
\textbf{Data:} Model architecture, network weights $W$, biases $b$, updated gradients of weights $\nabla W \ell$, updated gradients of biases $\nabla b \ell$ \\
\textbf{Result:} Reconstructed input $X^{(1)}$

\begin{algorithmic}[1]
\For{$i \leftarrow d$ to $1$} 
    \If{$i = d$} 
        \State $\nabla X^{(i)} \ell \leftarrow W^{(i)T} \cdot \nabla b^{(i)} \ell$
        \State $X^{(i)} \leftarrow \frac{\partial \ell}{\partial W^{(i)}_j} \bigg/ \frac{\partial \ell}{\partial b^{(i)}_j}$
    \Else 
        \State $\nabla Z^{(i)} \ell \leftarrow \nabla X^{(i+1)} \ell \cdot A'^{(i)}(Z^{(i)})$
        \State /*  $A'^{(i)}(Z^{(i)})$ can be derived directly using $X^{(i+1)}$ as detailed in Table 1 */

        \State $\nabla X^{(i)} \ell \leftarrow W^{(i)}_{\text{flipped}} * \nabla Z^{(i)} \ell$
        \State $\nabla W^{(i)} \ell \leftarrow \nabla Z^{(i)} \ell * X^{(i)}$
        \State $X^{(i)} \leftarrow \nabla Z^{(i)^-1} * \nabla W^{(i)} \ell$
    \EndIf
\EndFor

\State \textbf{return} $X^{(1)}$ \Comment{Reconstructed original input}
\end{algorithmic}
\end{algorithm}

\subsection{High-Resolution Image Reconstruction}
This algorithm extends the reconstruction approach from single training examples to high-dimensional images by employing a hybrid strategy that dynamically integrates gradient constraints and parameter constraints. While the traditional methods relying solely on parameter constraints are applicable for low-dimensional inputs, they face significant limitations when applied to high-dimensional data. Our proposed method combines gradient constraints and parameter constraints in a hybrid approach to overcome this challenge.\newline

The reconstruction of the input for the first fully connected layer employs the same methodology outlined in Equations (6) and (9) for the previous algorithm. This approach maintains robustness across varying input dimensionalities. The reconstruction process is formalized in Algorithm \ref{alg:s2}.\newline

The success of reconstructing the input from convolutional layers depends on two critical factors:
\begin{itemize}
    \item \textbf{The number of available equations:} Determines whether the system of linear equations can be solved accurately.
    \item \textbf{The dimensionality of the matrix to solve:} Impacts the computational complexity and feasibility of the reconstruction process.
\end{itemize}
The proposed algorithm employs different reconstruction methods for late convolutional layers, characterized by high-dimensional channels, and for early convolutional layers defined by high-dimensional inputs, to optimize computational efficiency.

\subsubsection{Reconstruction in Late Convolutional Layers with High-Dimensional Channels:}

In the deeper layers of a convolutional neural network, feature maps typically have reduced spatial dimensions due to striding operations, while the number of channels increases significantly. The Gradient Constraints method is particularly effective in this context, as it reconstructs each channel independently, maintaining manageable matrix sizes. The reconstruction process for late convolutional layers using gradient constraints is detailed in Equations (10) and (15) of the previous algorithm. The core advantage of this method lies in its ability to reconstruct channels separately, significantly reducing the dimensionality of the matrix that needs to be solved during reconstruction. The matrix dimension $(D_M)$ can be expressed in terms of the input feature map height $(Height)$, input feature map width $(Width)$, and number of channels $(Channels)$ as follows:

\begin{itemize}
    \item \textbf{Using Parameter Constraints:} When all channels must be reconstructed simultaneously, the matrix dimension is:

    \[
    D_M (\text{Parameter Constraints}) = Width \times Height \times Channels
    \]

    \item \textbf{Using Gradient Constraints:} When each channel can be reconstructed independently, the matrix dimension is significantly reduced to:

    \[
    D_M (\text{Gradient Constraints}) = Width \times Height
    \]
\end{itemize}

This reduction is crucial, as the number of channels in late convolutional layers is often significantly high, and managing a matrix of this size under weight constraints would impose substantial computational challenges. By breaking down the reconstruction into individual channels, the gradient constraints method ensures scalability and efficiency, making it an ideal choice for deep layers of the network.

\subsubsection{Reconstruction in Early Convolutional Layers with High-Dimensional Inputs:}

In the early convolutional layers, feature maps typically retain large spatial dimensions, while the number of channels remains relatively low. This characteristic necessitates a different approach to input reconstruction, as the high-dimensional input space imposes challenges when using gradient constraints.\newline 
The Gradient Constraints method requires a sufficient number of filters to reconstruct the input accurately. The required number of filters $(Filters)$ can be computed as:

\[
Filters = \frac{Width \times Height}{\text{Kernel Size}}
\]

Since in early layers the feature map width and height are typically high, this translates into a large number of required filters, rendering the gradient constraints method impractical for early convolutional layers.

When using \textbf{Parameter Constraints}, the reconstruction approach can be represented as follows:

\begin{itemize}
    \item \textbf{Backpropagation Through Activation Function:} 
    Before reconstructing the input, the intermediate output \( Z^{(i)} \) is computed by applying the inverse of the fully-invertible activation function to the input \( X^{(i+1)} \):

\begin{equation}
    {Z^{(i)}} = {A^{-1}(X^{(i+1)})}
    \tag{16}
\end{equation}
    \item \textbf{Reconstruction Using Parameter Constraints:}
    With the computed \( Z^{(i)} \), the input \( X^{(i)} \) can be reconstructed using weight constraints:

\begin{equation}
    {X^{(i)}} = {W}^{(i)^{-1}} * {Z^{(i)}} - {b^{(i)}}
    \tag{17}
\end{equation}

    where:

    \begin{itemize}
        \item \( \mathbf{W^{(i)}} \) is the weight matrix of the convolutional layer.
        \item \( \mathbf{Z^{(i)}} \) denotes the output before the activation function.
        \item \( \mathbf{b^{(i)}} \) is the bias vector of the layer.
    \end{itemize}
\end{itemize}

\begin{algorithm}[t]
\caption{Reconstruction of High-Dimensional Images}
\label{alg:s2}
\textbf{Data:} Model architecture, network weights $W$, biases $b$, updated gradients of weights $\nabla W \ell$, updated gradients of biases $\nabla b \ell$ \\
\textbf{Result:} Reconstructed input $X^{(1)}$
\begin{algorithmic}[1]

\State /* Step 1: Reconstruct Input for Fully Connected Layer */
\State Compute $X^{(d)}$ using Equations (6) and (9) from Algorithm 1

\State /* Step 2: Backpropagate through Convolutional Layers */
\For{$i = d-1$ to $1$} 
    \If{Layer $i$ is a late convolutional layer}
        \State Use Gradient Constraints to compute $\nabla Z^{(i)}$ and reconstruct $X^{(i)}$:
        \State $\nabla Z^{(i)} \ell \leftarrow \nabla X^{(i+1)} \ell \cdot A'^{(i)}(Z^{(i)})$ 
        \State $\nabla X^{(i)} \ell \leftarrow W^{(i)}_{\text{flipped}} * \nabla Z^{(i)} \ell$
        \State $\nabla W^{(i)} \leftarrow \nabla Z^{(i)} * X^{(i)}$
        \State $X^{(i)} \leftarrow \nabla Z^{(i)^{-1}} * \nabla W^{(i)}$
        \State /* Gradient Constraints: Each channel reconstructed independently */
    \Else
        \State Use Parameter Constraints to reconstruct $X^{(i)}$:
    \State Compute the intermediate output before activation by backpropagating through the fully-invertible activation function:
    \[
    Z^{(i)} = A^{-1}(X^{(i+1)})
    \]
    \State Reconstruct the input using parameter (weight) constraints:
    \[
    X^{(i)} = W^{(i)^{-1}} * Z^{(i)} - b^{(i)}
    \]
        \State /* Parameter Constraints: All channels reconstructed simultaneously */
    \EndIf
\EndFor

\State \textbf{return} $X^{(1)}$ \Comment{Reconstructed original input}
\end{algorithmic}

\end{algorithm}

\subsection{Reconstruction of Images from Mini-Batches}
In practical federated learning (FL) scenarios, clients typically compute and transmit gradient updates using \textit{mini-batches} rather than single training examples. Most existing gradient inversion methods—both optimization-based and analytical—focus on reconstructing individual inputs, which limits their applicability in realistic FL deployments. 

\textbf{Motivation.} Reconstructing training data from mini-batch gradients is significantly more challenging due to the aggregation of gradient contributions from multiple samples. 
Our objective is to develop an \textit{analytical framework} capable of reconstructing individual data samples from a shared mini-batch gradient update, filling a critical gap in the literature where such analytical approaches are largely unexplored.




The proposed algorithm targets the reconstruction of input images from mini-batch gradients in a neural network architecture consisting of convolutional layers followed by a fully connected layer indexed as \( d \), where a cross-entropy loss function is employed. Unlike single-example reconstruction approaches, this method extends the reconstruction process to handle mini-batch processing, allowing the recovery of multiple training samples simultaneously. The complete reconstruction process is formalized in Algorithm \ref{alg:s3}.

Assuming the network uses a softmax activation function at the output of the fully connected layer, the predicted probability \( p_{k,c} \) for class \( c \) and example \( k \) is defined as:

\[
p_{k,c} = \frac{e^{z_{k,c}^{(d)}}}{\sum_{j=1}^C e^{z_{k,j}^{(d)}}}
\]

where \( z_{k,c}^{(d)} \) represents the logit for class \( c \) and example \( k \) in the fully connected layer \( d \), and \( C \) is the total number of classes. The gradient of the loss function \( \ell \) with respect to the logit \( z_{k,c}^{(d)} \) is expressed as:

\begin{equation}
\frac{\partial \ell}{\partial z_{k,c}^{(d)}} = p_{k,c} - y_{k,c} \tag{18}
\end{equation}

where \( y_{k,c} \) is the ground truth label (one-hot encoded) for class \( c \) and example \( k \). Aggregating this gradient over the entire mini-batch of size \( B \) gives:

\begin{equation}
\frac{\partial \ell}{\partial z_c^{(d)}} = \frac{1}{B} \sum_{k=1}^{B} (p_{k,c} - y_{k,c}) \tag{19}
\end{equation}
Since the batch size \( B \) is typically much smaller than the number of classes \( C \), and given the sparsity of the softmax probabilities, most classes are predominantly influenced by only a single example in the mini-batch. Assuming that class \( c \) is dominated by example \( k \), this observation allows for the approximation:

\begin{equation}
\frac{\partial \ell}{\partial z_c^{(d)}} \approx \frac{1}{B}(p_{k,c} - y_{k,c}) \tag{20}
\end{equation}

The gradient with respect to the bias \( b_c^{(d)} \) of the fully connected layer for class \( c \) can be approximated by:

\begin{equation}
\frac{\partial \ell}{\partial b_c^{(d)}} = \frac{\partial \ell}{\partial z_c^{(d)}} \approx \frac{1}{B}(p_{k,c} - y_{k,c}) \tag{21}
\end{equation}

Similarly, the gradient with respect to the weights \( W_c^{(d)} \) for the same class is:

\begin{equation}
\frac{\partial \ell}{\partial W_c^{(d)}} = \frac{1}{B} \sum_{k=1}^{B} (p_{k,c} - y_{k,c}) X_k^{(d)} \approx \frac{1}{B}(p_{k,c} - y_{k,c}) X_k^{(d)} \tag{22}
\end{equation}

By dividing Equation (21) by Equation (20), the input vector \( X_k^{(d)} \) can be reconstructed as:

\begin{equation}
X_k^{(d)} = X_c^{(d)}= \frac{\partial \ell}{\partial W_c^{(d)}} \bigg / \frac{\partial \ell}{\partial b_c^{(d)}} \tag{23}
\end{equation}

Notably, a single output node (class) is sufficient to reconstruct an entire input vector. This characteristic enables the algorithm to generate as many input vectors as there are classes. Given that the number of classes is generally greater than the number of examples in the mini-batch, this approach maximizes the potential for complete reconstruction.

For each reconstructed input vector derived from the fully connected layer, the algorithm employs the same backpropagation approach illustrated in Equations (16) and (17) to iteratively propagate the input through the convolutional layers. By repeating this process for all input vectors, the method achieves full reconstruction of the input images in the mini-batch.

\begin{algorithm}[t]
\caption{Reconstruction of Images from Mini-Batches}
\label{alg:s3}
\textbf{Data:} Model architecture, network weights $W$, biases $b$, updated gradients of weights $\nabla W \ell$, updated gradients of biases $\nabla b \ell$
\\
\textbf{Output:} Reconstructed input images $\{X_k\}$ for examples in the mini-batch

\begin{algorithmic}[1]
\State /* Step 1: Construct Input Vectors for Fully Connected Layer */
\For{$c = 1$ to $C$} \Comment{C is the number of classes}
    \State Compute input vector $X_c^{(d)}$ for each class $c$ using:
    \[
    X_c^{(d)} = \frac{\partial \ell}{\partial W_c^{(d)}} \bigg / \frac{\partial \ell}{\partial b_c^{(d)}}
    \]
\EndFor

\State /* Step 2: Backpropagate Each Input Vector Through Convolutional Layers Using Weight Constraints */
\For{$k = 1$ to $C$} 
    \State Initialize $X^{(d)} \leftarrow X_c^{(d)}$ 
    \For{$i = d-1$ to $1$} 
            \State Backpropagate $X^{(d)}$ through the fully-invertible activation function to compute $Z^{(i)}$:
        \[
        Z^{(i)} = A^{-1}(X^{(i+1)})
        \]
        \[
        X^{(i)} = W^{(i)^{-1}} * Z^{(i)} - b^{(i)}
        \]
    \EndFor
    \State Store reconstructed input $X^{(i)}$ as $X_k$
\EndFor

\State \textbf{Return} $\{X_k\}$ \Comment{Reconstructed images examples in the mini-batch}
\end{algorithmic}
\end{algorithm}
The algorithm proceeds by:

\begin{itemize}
    \item Constructing Input Vectors: Generate input vectors for the fully connected layer by considering each class separately.
    \item Isolated Backpropagation: Backpropagate each input vector individually through the convolutional layers using the methods outlined in the previous algorithms for reconstructing single examples.
\end{itemize}

\section{Analysis of Recursive Gradient on Convolutional Layers}

In this section, we introduce an analytical framework to assess the feasibility of input reconstruction in convolutional layers using both gradient and weight constraints. This analysis enables us to determine the conditions under which recovery is possible, based on the network architecture, the characteristics of the activation function, and the relationship between the number of constraints and the input dimensionality.

During the forward pass of a convolutional layer \(i\), the intermediate output \(Z^i\) is computed as the convolution of the input \(X^i\) with the filter weights \(W^i\), represented by:

\begin{equation}
Z^i = W^i * X^i + b^i
\tag{24}
\end{equation}

where \( * \) denotes the convolution operation. The final output \(O^i\) of the layer is obtained by applying the activation function to \(Z^i\).

In the backward pass, the gradient with respect to the filter weights \( \nabla W^i \) is computed by convolving the input \( X^i \) with the gradient of the output \( \nabla Z^i \), expressed as:

\begin{equation}
\nabla W^i = \nabla Z^i * X^i
\tag{25}
\end{equation}

Combining the forward and backward pass equations leads to a unified system:

\begin{equation}
\begin{bmatrix}
W^i \\
\nabla Z^i
\end{bmatrix} * X^i = \begin{bmatrix}
Z^i \\
\nabla W^i
\end{bmatrix}
\tag{26}
\end{equation}

For successful reconstruction of the input \( X^i \), the rank of the matrix 
\(\begin{bmatrix} W^i \\ \nabla Z^i \end{bmatrix}\) must equal the number of unknowns \( |X^i| \).
The total number of constraints comprises weight constraints derived from the forward pass, influenced by the activation function's invertibility, and gradient constraints resulting from the backward pass, contingent on the ability to propagate gradients through the activation function.

The number of weight constraints depends on the reconstructible intermediate outputs \( |Z^i| \) from the subsequent layer \( i+1 \), varying with the activation function. For fully invertible activation functions, the constraints equal the total number of output data points, \( |W^i| = |Z^i| \); for non-invertible activation functions, \( |W^i| = 0 \); and for partially invertible activation functions, \( |W^i| \) forms a subset of \( |Z^i| \), proportional to the reconstructible data points.

Gradient constraints are viable when the gradients with respect to the layer's output \( \nabla Z^i \) are known, with \( |\nabla Z^i| = |W^i| \). The feasibility of deriving \( \nabla Z^i \) hinges on the successful propagation of gradients through the activation function, as detailed in the methodology. When feasible, the number of gradient constraints matches the number of weights in the layer.

The number of weight constraints correlates with the output dimension, while gradient constraints align with the total weights in the layer. At the initial layers of the network, the output dimensions are large, and the input channels are relatively few, leading to \( |W^i| > |\nabla Z^i| \). As the network depth increases, the spatial dimensions of feature maps typically reduce while the number of channels rises, resulting in \( |W^i| < |\nabla Z^i| \). This shift emphasizes the increased effectiveness of gradient constraints in deeper layers, contradicting prior assertions by~\cite{zhu2020r} that suggested minimal viable gradient constraints in convolutional layers.

To prevent gradient-based attacks, it is imperative to ensure that:

\[
|X^i| > |W^i| +|\nabla Z^i|
\]

Our analysis identifies the number of filters in the convolutional layer as a pivotal factor in enabling or mitigating analytical attacks. Quantitatively, given an input of dimension \( Height \times Width \times Channels \), with \( Filters \) convolutional filters, a kernel size \( K \), stride \( S \), and padding \( P \):

\begin{equation}
|W^i|= \left(\frac{Height+2P-K}{S}+1\right) \cdot Filters
\tag{27}
\end{equation}

\begin{equation}
|\nabla Z^i|= K^2 \cdot Filters
\tag{28}
\end{equation}

Adjusting these architectural parameters provides control over the weight and gradient constraints, directly impacting the attack's feasibility.

\section{Results and Discussions}
\label{Results}
\begin{figure}[!t]
     \centering
     \begin{subfigure}[b]{0.2\textwidth}
         \centering
         \includegraphics[width=\textwidth]{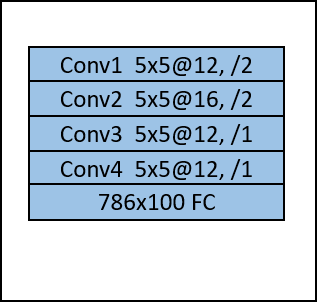}
         \caption{LeNet}
         \label{fig:Letnet}
     \end{subfigure}
     \hfill
     \begin{subfigure}[b]{0.2\textwidth}
         \centering
         \includegraphics[width=\textwidth]{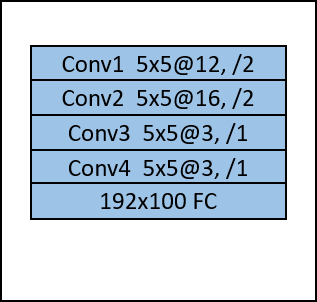}
         \caption{LeNet-x}
         \label{fig:letnet_O}
     \end{subfigure}
     \hfill
     \begin{subfigure}[b]{0.2\textwidth}
         \centering
         \includegraphics[width=\textwidth]{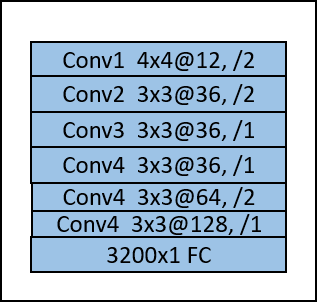}
         \caption{CNN6}
         \label{fig:CNN6}
     \end{subfigure}
          \hfill
     \begin{subfigure}[b]{0.2\textwidth}
         \centering
         \includegraphics[width=\textwidth]{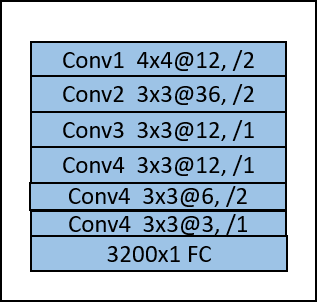}
         \caption{CNN6-x}
         \label{fig:CNN6-O}
     \end{subfigure}
          \begin{subfigure}[b]{0.2\textwidth}
         \centering
         \includegraphics[width=\textwidth]{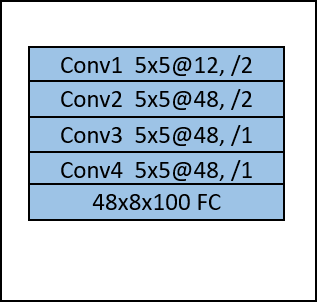}
         \caption{LeNet-ex}
         \label{fig:CNN6-O}
     \end{subfigure}
        \caption{The Five neural network architectures used for evaluation: (a) LeNet, as employed in the DLG baseline, representing a standard model configuration. (b) CNN6, utilized in the R-GAP baseline. (c) LeNet-x, a modified version of LeNet with a reduced number of filters, demonstrates our method's effectiveness in reconstructing images even with fewer parameters. (d) CNN6-x, an optimized version of CNN6 that reduces the number of filters in the last layer from 128 to 3, highlighting our approach’s robustness in scenarios with minimal parameterization. (e) LeNet-ex, an extended version of LetNet. The layer notation (e.g., “Conv1 5x5@12, /2”) specifies the kernel size (“5x5”), the number of filters (“12”), and the stride value (“/2”).}
        \label{fig:Different arrangements of layers}
\end{figure}

\textbf{Experimential Setup:} To evaluate the effectiveness of our proposed algorithms, we conducted extensive experiments on four datasets: CIFAR-10 \cite{krizhevsky2009learning}, CIFAR-100 \cite{krizhevsky2009learning}, MNIST \cite{lecun1998gradient}, and ImageNet \cite{5206848}, benchmarking our approach against two gradient inversion attack methods DLG \cite{zhu2019deep} and R-GAP \cite{zhu2020r}. The experiments aimed to assess the reconstruction performance under both standard and challenging conditions, demonstrating the superiority of our method in reconstructing input data from gradients under diverse settings. Our evaluation involved testing the proposed method on five distinct neural network architectures, including standard architectures used by baseline methods and modified versions with reduced filters. We measured performance using Mean Squared Error (MSE), Peak Signal-to-Noise Ratio (PSNR), and reconstruction time to ensure a comprehensive assessment of both accuracy and efficiency. The experiments were conducted on a 12th Gen Intel(R) i7 2.10 GHz processor with 32 GB RAM, showcasing the enhanced efficiency of our method in achieving faster reconstruction times than DLG and R-GAP. The implementation was carried out using the PyTorch framework. Our evaluation employed five CNN architectures, as depicted in Figure ~\ref{fig:Different arrangements of layers}: LeNet, CNN6, and their modified versions, LeNet-x, LetNet-ex, and CNN6-x:

\begin{itemize}
    \item CNN6: The architecture used in the R-GAP baseline.
        \item CNN6-x: A modified version of CNN6 with fewer convolutional filters, demonstrating the effectiveness of our method even in scenarios with reduced information.
    \item LeNet: The architecture employed in the DLG baseline.

    \item LeNet-x: A variant of LeNet with a reduced number of filters.
    \item LeNet-ex: An extended version of LeNet with an increased number of filters, to demonstrate the reconstruction process for high-dimensional images and mini-batch scenarios.
\end{itemize}

Our proposed method, R-CONV, introduces a highly efficient technique for reconstructing the inputs of convolutional layers using minimal parameters. The experimental results demonstrate that R-CONV not only accelerates image reconstruction compared to both DLG and R-GAP but also consistently generates higher-quality images. Unlike traditional approaches, our method is compatible with a broad range of activation functions, including those that are not fully invertible—overcoming a critical limitation of existing methods.

To validate our approach, we compared R-CONV with DLG using the LeNet and LeNet-x architectures and benchmarked it against R-GAP using CNN6 and CNN6-x architectures. Table \ref{tab:quant_results} presents a comprehensive quantitative analysis of our method's performance in terms of MSE, PSNR, and reconstruction time, offering a direct comparison with DLG and R-GAP. 

\begin{figure}[!ht]
     \centering
     \begin{subfigure}[b]{0.45\textwidth}
         \centering
         \includegraphics[width=\textwidth]{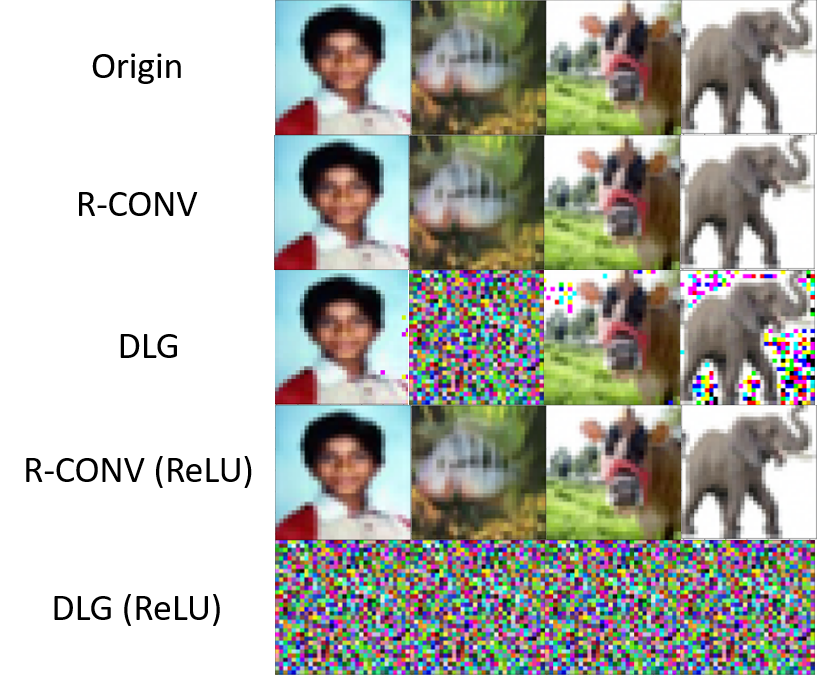}
         \caption{CIFAR100}
         \label{fig:Cifar10}
     \end{subfigure}
     \hfill
     \begin{subfigure}[b]{0.45\textwidth}
         \centering
         \includegraphics[width=\textwidth]{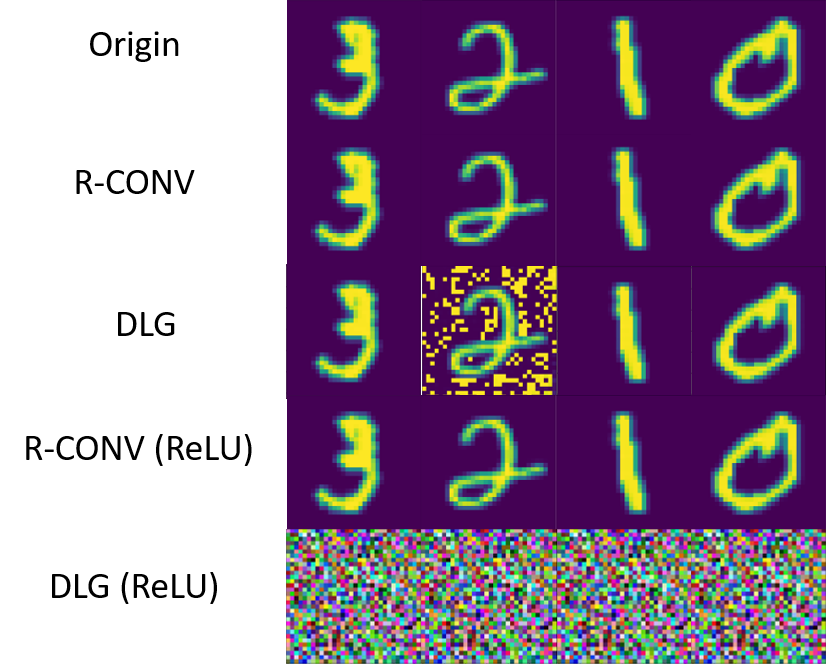}
         \caption{MNIST}
         \label{fig:MNIST}
     \end{subfigure}

        \caption{Reconstruction Quality Comparison Between R-CONV and DLG: Subfigure (a) presents reconstruction results using images from the CIFAR-100 dataset, while subfigure (b) illustrates reconstruction results with images from the MNIST dataset.}
        \label{fig:Reconstruction_Quality_CIFAR}
\end{figure}

\begin{figure}[!ht]
     \centering
         \includegraphics[width=\textwidth]{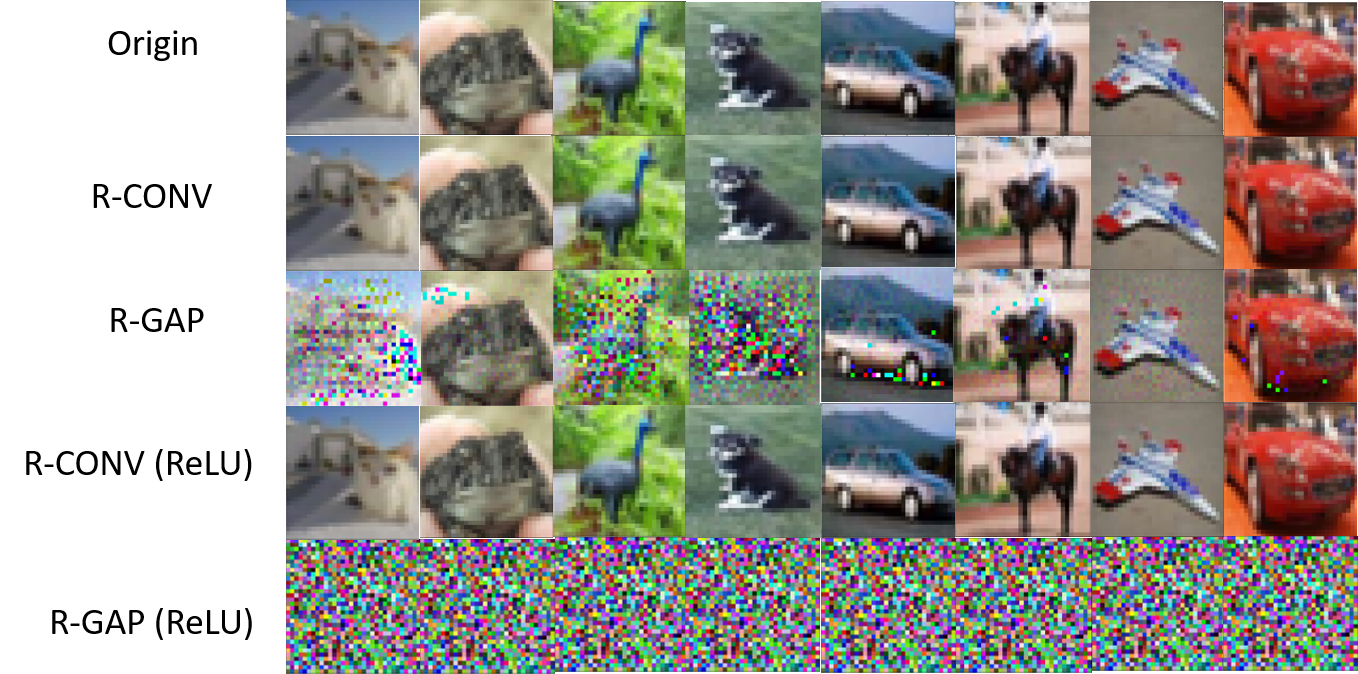}
        \caption{Reconstruction Quality Comparison Between R-CONV and R-GAP: This figure showcases the reconstruction results of images from the CIFAR-10 dataset. The first row displays the original images, followed by the second row showing reconstructions by R-CONV using the CNN6-O architecture. The third row presents reconstructions by R-GAP with the CNN6 architecture. The fourth row illustrates the reconstruction performance of R-CONV with ReLU activation, while the fifth row demonstrates the results of R-GAP under the same activation setting. The comparison highlights the superior performance of R-CONV, maintaining high-quality reconstructions even with ReLU activation, in contrast to R-GAP.}
        \label{Reconstruction_Quality_CIFAR}
\end{figure}

The results highlight R-CONV's remarkable efficiency in achieving high-quality reconstructions with fewer parameters. Importantly, while DLG’s performance significantly drops when the activation function is switched from Sigmoid to ReLU, R-CONV maintains its reconstruction quality, showcasing its robustness across different activation functions. Additionally, the results indicate a substantial improvement in reconstruction speed compared to R-GAP. This speedup is attributed to R-CONV's ability to fully leverage gradient constraints, allowing for a more reliable and accurate reconstruction process without the need to reconstruct the output of each layer—especially critical in scenarios where non-invertible activation functions are present.
Figure \ref{fig:Reconstruction_Quality_CIFAR} showcases the qualitative reconstruction performance on CIFAR-100 and MNIST datasets, highlighting the superior quality of images reconstructed using R-CONV. Our method consistently produces high-quality reconstructions, even under challenging activation functions where other methods fail. For comparisons with R-GAP, we utilized the CNN6 and CNN6-x architectures, evaluating performance on the CIFAR-10 dataset. As depicted in Figure \ref{Reconstruction_Quality_CIFAR}, R-CONV not only outperforms R-GAP in reconstruction quality but also demonstrates robustness with non-fully invertible activation functions, a critical limitation of R-GAP.


\begin{table}[!t]
\footnotesize
\centering
\caption{Comparison of the proposed R-CONV method with state-of-the-art analytical (R-GAP) and optimization-based (DLG) methods in terms of average MSE, PSNR, and reconstruction time. \textit{Our proposed method outperforms these state-of-the-art methods in all the considered metrics.}}
\label{tab:quant_results}

\begin{tabular}{cccc}
\toprule
 Method & MSE                                            & PSNR (dB) & Time (s) \\ \midrule 
\multicolumn{4}{c}{Average Computed for Images Depicted in Figure \ref{fig:Reconstruction_Quality_CIFAR}.}  \\ \midrule
 R-CONV  & $2.2 \times 10^{-7} \pm 3.64 \times 10^{-9}$  & $114.68\pm 5.5$ & $6.33 \pm 2.3$   \\
 DLG     & $0.0933\pm 0.05$                                       & $62.62 \pm 8.5$  & $60.66 \pm 5.58$  \\ \midrule \midrule
\multicolumn{4}{c}{Average Computed for Images Depicted in Figure \ref{Reconstruction_Quality_CIFAR}.} \\ \midrule
R-CONV  & $2.88 \times 10^{-9} \pm 2.44 \times 10^{-10}$& $150.12 \pm 4.5$ & $2.494 \pm 1.66$  \\
R-GAP   & $0.0056 \pm 0.008$                                       & $76.73 \pm 6.88$  & $232.45\pm 12.44$ \\ \bottomrule
\end{tabular}
\end{table}

\begin{figure}[!ht]
     \centering
     \begin{subfigure}[b]{0.45\textwidth}
         \centering
         \includegraphics[width=\textwidth]{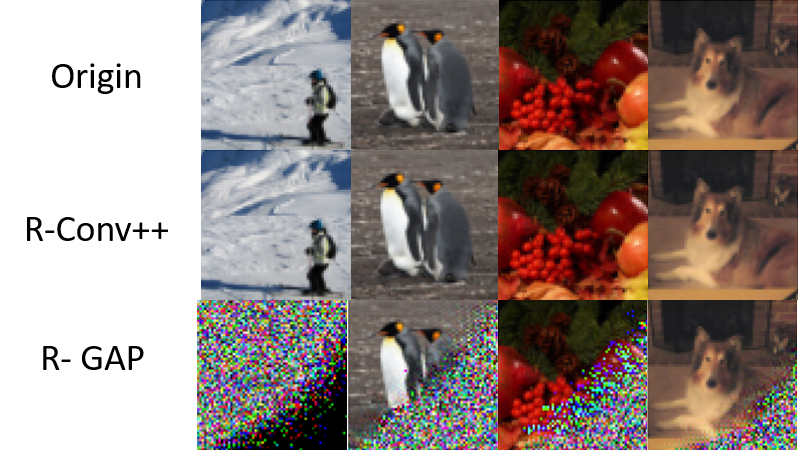}
         \caption{64 X 64}
         \label{fig:64}
     \end{subfigure}
     \hfill
     \begin{subfigure}[b]{0.45\textwidth}
         \centering
         \includegraphics[width=\textwidth]{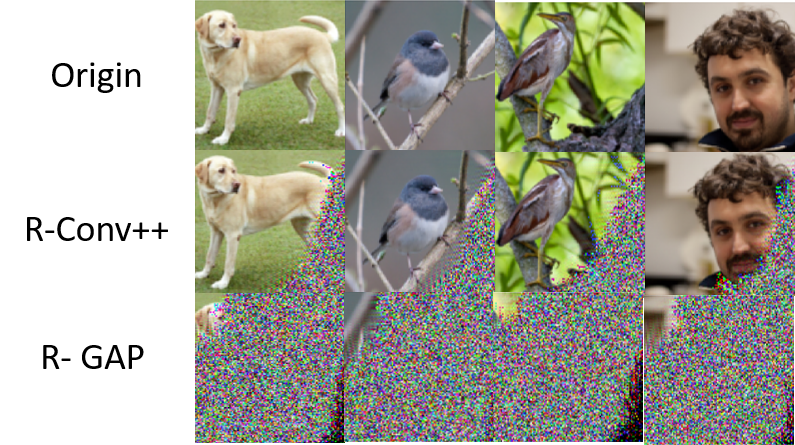}
         \caption{112 X 122}
         \label{fig:112}
     \end{subfigure}

        \caption{High-dimensional image reconstruction results on the ImageNet dataset using the LeNet-ex architecture. (a) Reconstruction performance at 64 × 64 resolution shows near-perfect results by our hybrid method compared to R-GAP. (b) At 112 × 112 resolution, while some degradation is observed, our approach maintains superior reconstruction quality over R-GAP, demonstrating robustness in handling high-dimensional data.}
        \label{fig:High}
\end{figure}
\begin{figure}[!ht]
     \centering
         \includegraphics[width=\textwidth]{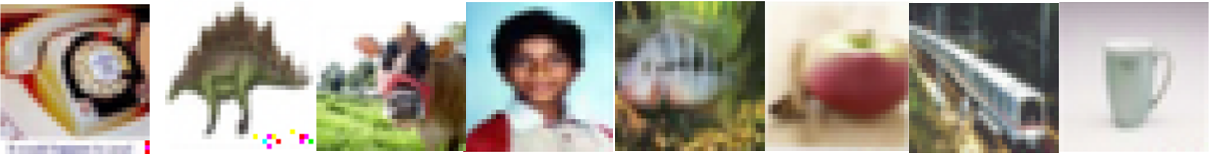}
        \caption{Reconstruction results with a mini-batch size of 8: All 8 images were successfully reconstructed.}
        \label{Reconstruction8}
\end{figure}
\begin{figure}[!ht]
     \centering
         \includegraphics[width=\textwidth]{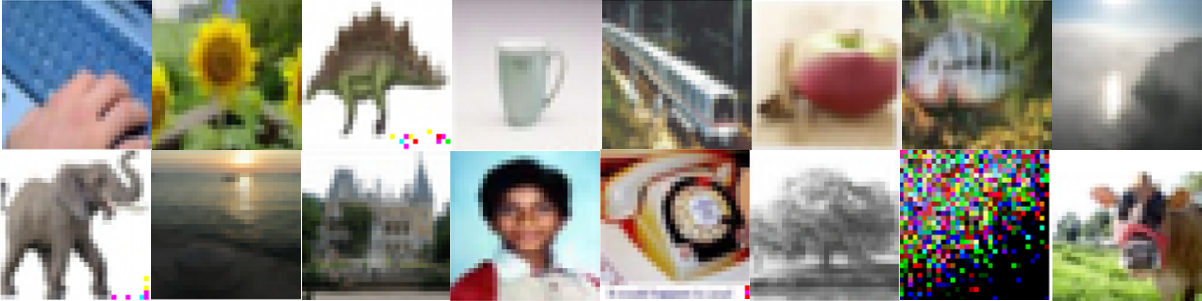}
        \caption{Reconstruction results with a mini-batch size of 16: 15 out of 16 images were successfully reconstructed.}
        \label{Reconstruction16}
\end{figure}
\begin{figure}[!ht]
     \centering
         \includegraphics[width=\textwidth]{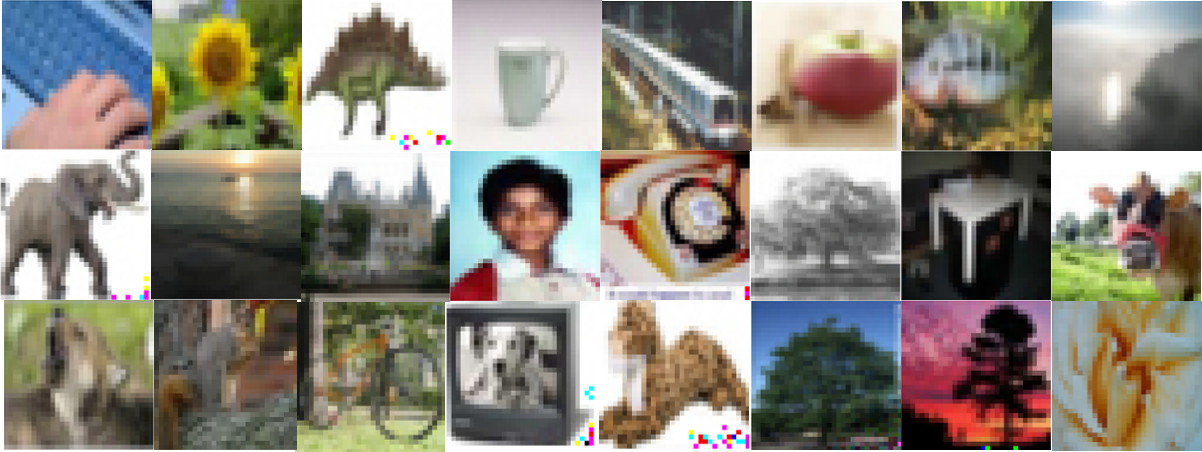}
        \caption{Reconstruction results with a mini-batch size of 32: 24 out of 32 images were successfully reconstructed.}
        \label{Reconstruction32}
\end{figure}
\begin{figure}[!ht]
     \centering
         \includegraphics[width=\textwidth]{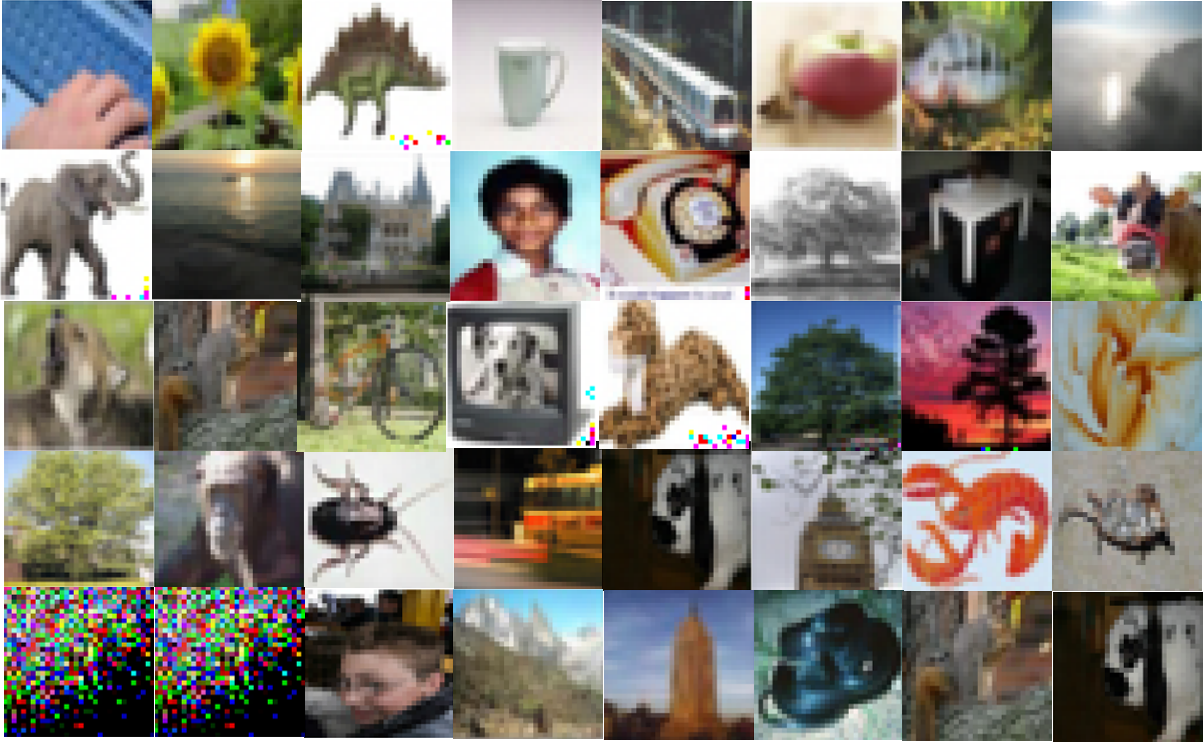}
        \caption{Reconstruction results with a mini-batch size of 48: 32 out of 48 images were successfully reconstructed.}
        \label{Reconstruction48}
\end{figure}
\subsubsection{High-Dimensional Image Reconstruction Using the Hybrid Approach}

To evaluate the effectiveness of our hybrid approach in reconstructing high-dimensional data, we extended the image resolution from $32 \times 32$ to $64 \times 64$ and further to $112 \times 112$. The experiments were conducted using the ImageNet dataset with the LeNet-ex architecture. Our hybrid method, which combines Gradient Constraints for the reconstruction of the last two convolutional layers and Parameter Constraints for the first two convolutional layers, was compared against the R-GAP method, which solely relies on parameter constraints.

Reconstruction Quality at Different Resolutions:

\begin{itemize}
    \item 64 × 64 Resolution: As shown in Figure \ref{fig:High}a, our hybrid approach achieved near-perfect reconstruction quality, significantly outperforming the R-GAP method. The gradient constraints employed in the deeper convolutional layers proved effective in maintaining reconstruction fidelity, whereas R-GAP struggled to preserve image details.
    
    \item 112 × 112 Resolution: When scaling up to 112 × 112, Figure \ref{fig:High}b illustrates that while our method experienced some degradation in reconstruction quality, it still maintained a considerable advantage over R-GAP. The weight constraints utilized in the early convolutional layers of our approach facilitated the handling of the high-dimensional input, and the adaptive transition to gradient constraints for deeper layers contributed to a balanced and effective reconstruction.
\end{itemize}

These results highlight the strength of our hybrid method in managing the challenges posed by high-dimensional image reconstruction. The ability to dynamically switch between gradient and parameter constraints ensures robust performance even as input dimensions increase, demonstrating a significant improvement over traditional methods that rely exclusively on parameter constraints.

\subsubsection{Reconstruction of Images from Mini-Batches}

To evaluate the performance of our method in reconstructing images from mini-batch gradients, we conducted experiments using the LeNet-ex architecture on the CIFAR-100 dataset, which comprises 100 classes. We tested the robustness and scalability of our approach with varying mini-batch sizes of 8, 16, 32, and 48, analyzing the algorithm's effectiveness in reconstructing input images under different settings.

The experimental results demonstrate that our method effectively reconstructs the majority of images within the mini-batch, even as the batch size increases. When the batch size was set to 8, our approach achieved a 100\% reconstruction rate, successfully reconstructing all 8 images (see Figure~\ref{Reconstruction8}). For a batch size of 16, our method reconstructed 15 out of 16 images, showing only a slight drop in performance (see Figure~\ref{Reconstruction16}). As the batch size increased to 32, our approach managed to reconstruct 24 images (see Figure~\ref{Reconstruction32}), and for the batch size of 48, it reconstructed 32 images (see Figure~\ref{Reconstruction48}).
\subsection{Defense Mechanisms:} 
Since our proposed approach reveals vulnerabilities in gradient sharing by enabling input reconstruction, it is crucial to consider defense mechanisms to mitigate such attacks. On the algorithmic level, one common defense is gradient perturbation, which introduces differential privacy by adding noise to gradients. While effective, this method requires careful calibration to avoid degrading model utility. Gradient sparsification or compression techniques can also obscure sensitive information by limiting the granularity of shared gradients. Additionally, secure aggregation protocols prevent the server from accessing individual gradients in plaintext, thereby obstructing direct reconstruction attacks. 

On the network design level, incorporating certain architectural components can inherently reduce the efficacy of reconstruction attacks. For example, using composite activation functions such as GELU or SiLU introduces non-linearities that complicate gradient inversion, especially in deeper networks. Furthermore, pooling layers, particularly max-pooling, reduce spatial resolution and discard local information, posing additional input recovery challenges. A combined approach that integrates both algorithmic and architectural defenses may offer robust privacy protection in distributed learning settings.

\subsection{Limitations:}
Our proposed R-CONV method and extensions for high-dimensional data reconstruction and mini-batch processing have demonstrated strong performance in generating high-quality image reconstructions. However, certain limitations should be acknowledged. One key limitation lies in handling composite activation functions such as GELU, SiLU, and Swish, where R-CONV's effectiveness in exploiting gradient constraints diminishes due to the inability to express their derivatives solely in terms of their outputs. Additionally, while our method effectively reconstructs high-dimensional images and supports mini-batch processing, these advanced capabilities are currently constrained to scenarios where fully invertible activation functions are employed, limiting applicability in more complex network architectures.




\section{Conclusion}
In this paper, we introduced a comprehensive framework for advancing gradient inversion attacks on shared model updates within the federated learning paradigm. Building upon our previous work, we proposed R-CONV++, an analytical method that efficiently reconstructs training data from the gradients of convolutional layers, emphasizing the critical role of gradient constraints over traditional weight constraints. Our extended methodology introduced two novel algorithms: one for high-dimensional image reconstruction using a hybrid strategy that integrates gradient and parameter constraints, and another for reconstructing images from mini-batch gradients by leveraging output node sparsity. 

Experimental results demonstrated the superiority of our approach compared to existing methods, such as DLG and R-GAP, in terms of reconstruction quality, computational efficiency, and versatility across a wide range of activation functions. Our approach maintains robustness even with fewer parameters and under non-fully invertible activation functions, highlighting its applicability to real-world scenarios. 



\section*{Funding}

This study is supported by:
\begin{itemize}
    \item Qatar National Research Fund, Grant Number: NPRP12C‐33905‐SP‐66.
    \item the "COLTRANE-V" project – funded by the Ministero dell’Università e della Ricerca – within the PRIN 2022 program (D.D.104 - 02/02/2022). 
    \item SERICS project (PE00000014) under the MUR National Recovery and Resilience Plan funded by the European Union - NextGenerationEU. 
    \item  IRCC-2024-499 International Research Collaboration Co-Fund between Qatar University, Paris-Dauphine and Madrid Universities
\end{itemize}

This manuscript reflects only the authors’ views and opinions, and the funding bodies cannot be considered responsible for them.

\bibliographystyle{splncs04}
\bibliography{ref}
\end{document}